\documentclass[useamsfonts]{pasj01}

\usepackage{graphicx,todonotes}
\usepackage{color}
\usepackage{aas_macros}
\usepackage{amssymb}
\usepackage[fleqn]{amsmath}
\usepackage{natbib}
\newcommand{\simgt}{\lower.5ex\hbox{$\; \buildrel > \over \sim \;$}}
\newcommand{\simlt}{\lower.5ex\hbox{$\; \buildrel < \over \sim \;$}}

\def\mvir{\mbox{$M_\mathrm{200c}$}}
\def\mwl{\mbox{$M_\mathrm{WL}$}}
\def\msz{\mbox{$M_\mathrm{SZ}$}}
\def\mhunit{\mbox{$10^{14}\,M_\odot/h$}}
\def\munit{\mbox{$10^{14}\,M_\odot$}}
\def\cvir{\mbox{$c_\mathrm{200c}$}}

\def\mpch{\mbox{$\mathrm{Mpc}/h$}}

\def\sig8{\mbox{$\sigma_8$}}
\def\Sigmacr{\mbox{$\Sigma_{\rm cr}$}}
\def\DSigma{\mbox{$\Delta\Sigma$}}

\def\equationautorefname~#1\null{Equation~(#1)\null}

\begin{document}
\title{{\it Planck } Sunyaev-Zel'dovich Cluster Mass Calibration using Hyper Suprime-Cam Weak Lensing} 

\author{Elinor {Medezinski}\altaffilmark{1}}
\author{Nicholas {Battaglia}\altaffilmark{1}}
\author{Keiichi {Umetsu}\altaffilmark{2}}
\author{Masamune {Oguri}\altaffilmark{3,4,5}}
\author{Hironao {Miyatake}\altaffilmark{3,6}}
\author{Atsushi J. {Nishizawa}\altaffilmark{7}}
\author{Crist\'obal {Sif\'on}\altaffilmark{1}}
\author{David N. {Spergel}\altaffilmark{1,8}}
\author{I-Non {Chiu}\altaffilmark{2}}
\author{Yen-Ting {Lin}\altaffilmark{2}}
\author{Neta {Bahcall}\altaffilmark{1}}
\author{Yutaka {Komiyama}\altaffilmark{9,10}}

\email{elinorm@astro.princeton.edu}

\altaffiltext{1}{Department of Astrophysical Sciences, Princeton University, Princeton, NJ 08544, USA} 
\altaffiltext{2}{Institute of Astronomy and Astrophysics, Academia
Sinica, P.~O. Box 23-141, Taipei 10617, Taiwan.}  
\altaffiltext{3}{Kavli Institute for the Physics and Mathematics of the Universe (Kavli IPMU, WPI), TokyoInstitutes for Advanced Study, The University of Tokyo, Chiba 277-8582, Japan}
\altaffiltext{4}{Research Center for the Early Universe, University of Tokyo, Tokyo 113-0033, Japan}
\altaffiltext{5}{Department of Physics, University of Tokyo, Tokyo 113-0033, Japan}
\altaffiltext{6}{Jet Propulsion Laboratory, California Institute of Technology, Pasadena, CA 91109, USA}
\altaffiltext{7}{Institute for Advanced Research, Nagoya University, Nagoya 464-8602, Aichi, Japan}
\altaffiltext{8}{Center for Computational Astrophysics, Flatiron Institute, 162 5th Ave. New York, NY 10010}
\altaffiltext{9}{National Astronomical Observatory of Japan, 2-21-1 Osawa, Mitaka, Tokyo 181-8588, Japan}
\altaffiltext{10}{Department of Astronomy, School of Science, SOKENDAI (The Graduate University for Advanced Studies), 2-21-1 Osawa, Mitaka,  Tokyo 181-8588, Japan}


\maketitle

\begin{abstract}
Using $\sim$140 deg$^2$ Subaru Hyper Suprime-Cam
(HSC) survey data, we stack the  weak lensing (WL) signal  around five  {\it Planck} clusters found
within the footprint. This yields a 15$\sigma$ detection of the mean {\it Planck}
cluster mass density profile. The five {\it Planck} clusters span a relatively wide mass range, $M_{\rm WL,500c} = (2-30)\times\munit$ with a mean mass of $M_{\rm WL,500c} = (4.15\pm0.61)\times\munit$. The  ratio of the stacked {\it Planck} Sunyaev-Zel'dovich (SZ) mass to the stacked WL mass is $ \langle\msz\rangle/\langle\mwl\rangle = 1-b = 0.80\pm0.14$. This mass bias is consistent with previous WL mass calibrations of {\it Planck} clusters within the errors.  
We discuss the implications of our findings for the calibration of SZ cluster counts and the much discussed tension between {\it Planck} SZ cluster counts and {\it Planck} $\Lambda$CDM cosmology.
\end{abstract}

\section{Introduction} 
\label{sec:intro}

The abundance of galaxy clusters, particularly at high redshifts, is sensitive to the cosmological parameters that describe structure formation such as the matter density ($\Omega_\mathrm{M}$) and  the normalization of the matter power spectrum \citep[\sig8, e.g.,][]{Bahcall1998,Henry2000,Henry2009,Reiprich2002,Voit2005,Allen2011}. 
Since Abell's seminal work \citep{Abell1958}, many ongoing efforts have yielded  detections of thousands of clusters \citep[e.g.,][]{Gladders2000,Wen2012,Rykoff2014,Vikhlinin2009,Mantz2010}.
In particular, the {\it Planck} satellite has provided an important catalog of over a thousand galaxy clusters to higher redshift \citep{Planck-Collaboration2014b,Planck-Collaboration2016}   through the thermal Sunyaev-Zel'dovich \citep[SZ;][]{Sunyaev1972} selection. 
Two other ongoing SZ surveys, namely the Atacama Cosmology Telescope \citep[ACT;][]{ACT2003}, the South Pole Telescope \citep[SPT;][]{Carlstrom2011} and their successors, keep pushing the detection limits to lower masses and higher redshifts \citep[e.g.,][]{Staniszewski2009,Marriage2011,Reichardt2013,Hasselfield2013,Bleem2015}. 

The mass observable in all these SZ experiments is the volume-integrated Intra-Cluster Medium (ICM) pressure, referred to as the Compton-$Y$ parameter. Since this SZ  proxy is not a direct probe of mass, scaling relations are typically invoked to translate it to a total cluster mass. In particular, {\it Planck} adopted a method that relies on X-ray observations of clusters to calibrate the $Y$ parameter \citep[e.g.,][]{Planck-Collaboration2014b}.  The initial X-ray determination of the total cluster mass, on the other hand, assumes the clusters are in hydrostatic equilibrium \citep[HSE,][]{EP2005,MA2007}.

The  cosmological  constraints placed by {\it Planck} SZ cluster counts have unveiled a modest tension between  $\sig8$ compared to constraints derived by combining the primary  cosmic microwave background (CMB) anisotropies with non-cluster data \citep[][]{Planck-Collaboration2014b,Planck-Collaboration2016}. If exacerbated by future data, this tension could be a signature of new physics, such as a larger non-minimal sum of neutrino masses, or more likely  could point to systematics  in the cluster mass calibration. The {\it Planck }SZ mass calibration relied on X-ray observations from XMM-Newton. 
There are a number of potential biases in this calibration. Clusters undergo mergers which would violate the assumption of HSE and result in few tens of percent bias  \citep{Rasia2006,Nagai2007a,Lau2009,Battaglia2012,Nelson2014,Henson2017}. In addition, several papers \citep{Mahdavi2013,Donahue2014,Rozo2014} argue there are  instrument calibration issues with XMM-Newton.

\begin{table*}
\tbl{{\it Planck }clusters  within HSC-Wide}{%
\begin{tabular}{clrrr}
\hline\hline
{\it Planck} Name & NED Name& R.A.$^1$ & Dec.$^1$ & $z$  \\
 & & {[deg]} & {[deg]}&   \\
\hline
PSZ2 G068.61-46.60 & Abell2457 & 338.91999 & 1.48489 & 0.0594 \\
PSZ2 G167.98-59.95 &Abell0329 & 33.67122 & $-$4.56735 & 0.1393 \\
PSZ2 G174.40-57.33& Abell0362 & 37.92156 & $-$4.88258 & 0.1843\\
PSZ2 G228.50+34.95& MaxBCGJ140.53188+03.76632 & 140.54565 & 3.77820 & 0.2701 \\
PSZ2 G231.79+31.48& MACSJ0916.1-0023/Abell0776 & 139.03851 & $-$0.40453 & 0.3324  \\
\hline
\end{tabular} }
\label{tab:clusters}
\begin{tabnote}
$^1$ BCG center (J2000).
\end{tabnote}
\end{table*}

Weak lensing (WL) offers an independent method for measuring and calibrating cluster masses, as it probes the total mass, regardless of the nature or dynamical state of this mass. Cluster weak lensing has matured significantly in the last two decades \citep[see reviews by][]{Bartelmann2001,Refregier2003,Hoekstra2008}. Dedicated  cluster simulations  have explored different aspects of systematics in the cluster mass derivation, including  the triaxiality of clusters \citep{Oguri2005,Corless2007,Becker2011},  the inclusion of line-of-sight structures \citep{Hoekstra2013}, the deviations of clusters from commonly adopted spherical \citep{NFW97} halos in the model extraction of total mass, and
the impact of baryonic effects \citep{Henson2017}. Proper source selection and photometric redshift biases  have been  explored in several pointed cluster WL  studies \citep{Medezinski2010,Okabe2010,Applegate2014,Gruen2017}. In parallel, image simulations have been utilized to robustly calibrate biases in galaxy shape measurements \citep{Heymans2006,Massey2007,Bridle2010,Kitching2012,Mandelbaum2015,Fenech-Conti2017}.

Several recent studies have used WL to recalibrate {\it Planck} SZ cluster masses, e.g., Weighing the Giants (WtG, 
\citealt{von-der-Linden2014}), the Canadian Cluster Cosmology Project (CCCP, \citealt{Hoekstra2015}), the Cluster Lensing And supernovae Survey with Hubble (CLASH, \citealt{Penna-Lima2016}; see also \citealt{Umetsu2014}; \citealt{Merten2015}),  and the Local Cluster Substructure Survey (LoCuSS; \citealt{Smith2016}).
These papers introduced a bias parameter to calibrate the measured SZ mass estimate with the true mass,
\begin{equation}
1-b \equiv \msz/M_{\rm True}.
\end{equation}
where \msz~ is the SZ mass estimate and $M_{\rm True}$ is the true mass. The best estimator for the true mass is assumed to come from weak lensing, \mwl~\citep[for caveats, see][]{Becker2011}. 
If the bias were zero ($b=0$), the {\it Planck} primary  CMB would predict far more clusters than observed. Reconciling  the  {\it Planck} SZ cluster counts requires $1-b=0.58$, about $2\sigma$ away from {\it Planck} adopted value, $1-b=0.8$ \citep{Planck-Collaboration2016}.
The WL mass calibrations differ in their conclusion as to what the bias level is, with some studies agreeing with the {\it Planck } value \citep[$b=0.3$--$0.4$;][]{von-der-Linden2014,Hoekstra2015,Penna-Lima2016,Sereno2017a}, and some finding little to no bias \citep[$b=0.1$--$0.2$;][]{Smith2016}. We note that the samples of clusters used in these studies  have marginal overlap in  redshift and mass ranges. It is not clear whether these differences are the result of systematics in the WL observations or that $1-b$ has a mass  or redshift dependence \citep{Andreon2014,Smith2016,Sereno2017b}.

In this paper, we address this important issue by calibrating the SZ masses of {\it Planck }clusters located within the latest WL observations of the  Hyper Suprime-Cam  Subaru Strategic Program \citep[HSC-SSP; see][]{Aihara2017,Aihara2017a}.
The HSC-SSP is an ongoing wide-field optical imaging survey with the HSC camera which is installed on the Subaru 8.2m telescope. Its Wide layer will observe the total sky area of $\sim1400$ deg$^2$ to $i\lesssim26$. With its unique combination of area and depth, the HSC Wide layer will both detect and provide accurate WL measurements of thousands of clusters to $z\sim1.5$. In its current stage, $\sim240$~deg$^2$ have been observed, out of which we use $\sim140$~deg$^2$ of full-depth and full-color (FDFC) data to characterize the five overlapping  {\it Planck } clusters and provide an independent measure of the SZ-WL mass ratio.

This paper is organized as follows. In Section~\ref{sec:data} we present the HSC survey and the {\it Planck }clusters found within HSC. In Section~\ref{sec:WLmethod} we describe the WL methodology. In Section~\ref{sec:results} we present the WL analysis and results, describing the source selection, the stacked and individual cluster WL analysis, the modeling, and the WL-SZ mass calibration. We summarize and conclude in Section~\ref{sec:summary}. Throughout this paper we adopt a {\it Wilkinson Microwave Anisotropy Probe} nine-year cosmology ({\it WMAP9}) \citep{Hinshaw2013}, where $\Omega_M=0.282,\ \Omega_\Lambda=0.718$, and $h=H_0/100$ km~s$^{-1}$ Mpc$^{-1} $.

\section{Data}
\label{sec:data}
\subsection{HSC observations}
\label{subsec:HSC}

The HSC-SSP \citep{Aihara2017a} is an optical imaging survey with the new HSC camera  (Miyazaki et  al., in prep) installed on the Subaru  8m telescope. The  HSC-SSP  survey consists of three layers: Wide, Deep and Ultradeep. The survey has been allocated 300 nights spanning five years (2014--2019). The Wide survey, when complete, will observe  $\sim$1400 deg$^2$.
In this study, we  use  the  current internal data release (S16A). It contains $\sim$140 deg$^2$ of FDFC area.    \cite{Aihara2017,Aihara2017a} give an overview of the survey and its public data release (S15B). The HSC Pipeline, \texttt{hscPipe}  \citep{Bosch2017}, based on the Large Synoptic Survey Telescope (LSST) pipeline \citep{LSST,Axelrod2010,Juric2015}, is used to reduce HSC data.

HSC-Wide consists of observations in five board-band filters, $grizy$, reaching a typical limiting magnitude of $i\simeq26$. So far it has reached exceptional  seeing with a median of FWHM$=0.6\arcsec$ in the $i$ band. Seven different codes have been employed by the team to produce  photometric redshift (photo-z) catalogs from the multi-band data \citep{Tanaka2017}. Here we make use of the {\sc mlz} photo-z code. 
Each galaxy is assigned a probability distribution function (PDF), from which various photo-z point estimates are derived (e.g., mean, median, etc.). We make use of the full PDF to avoid any potential biases of point estimators.

The WL shapes are estimated on the coadded $i$-band images using the re-Gaussianization method 
\citep{Hirata2003}, and are fully described in \cite{Mandelbaum2017}. Basic  cuts have been applied to these catalogs to ensure galaxies with robust  photometry and shapes. Further  photo-z quality cuts \citep{Tanaka2017} are  applied to the catalogs so that only galaxies with measured photo-z's remain. Cuts needed to obtain the source catalog are described in Section~\ref{subsec:CCsel}.

\subsection{{\it Planck }Cluster Sample in HSC}
\label{subsec:planck}

\begin{figure*}
\includegraphics[width=\textwidth,clip]{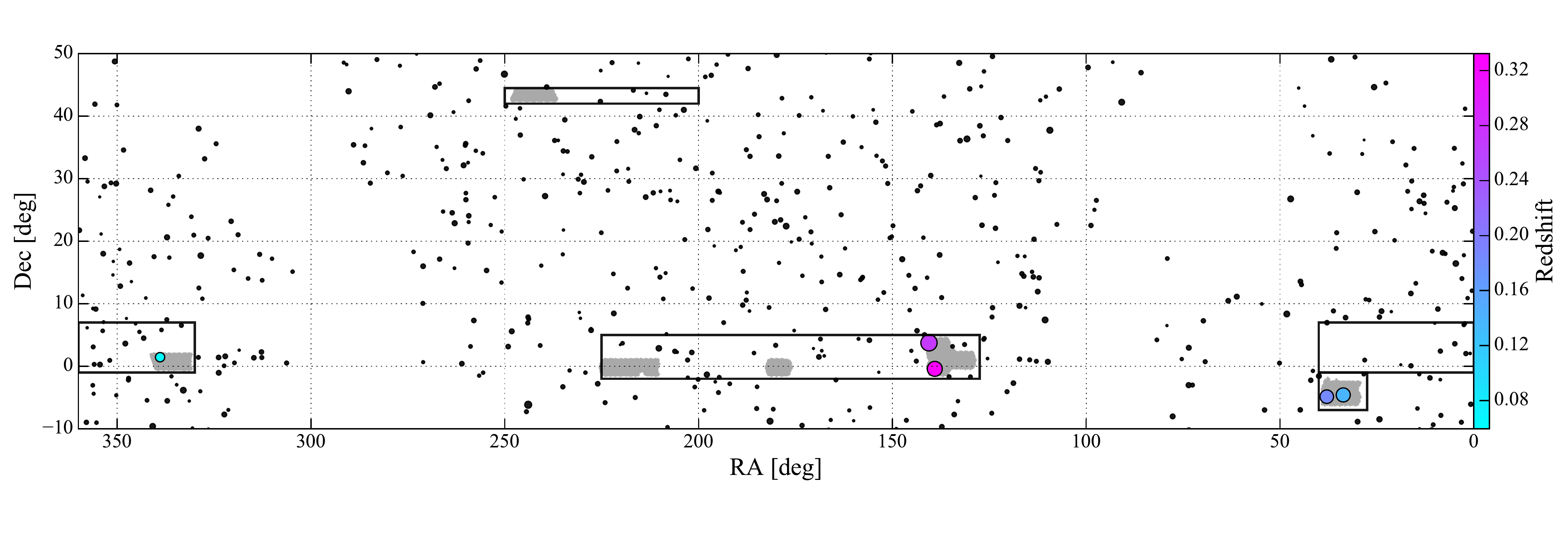}
\caption{Sky distribution of the HSC-SSP Wide fields (black outline) and the area observed thus far in FDFC (gray regions). Black circles are {\it Planck}-detected clusters, colored circles are the five {\it Planck }clusters within the HSC FDFC footprint. Color represents redshift, and circle size represents SZ mass.}
\label{fig:sky}
\end{figure*}

\begin{figure*}[t]
\includegraphics[width=\textwidth,clip]{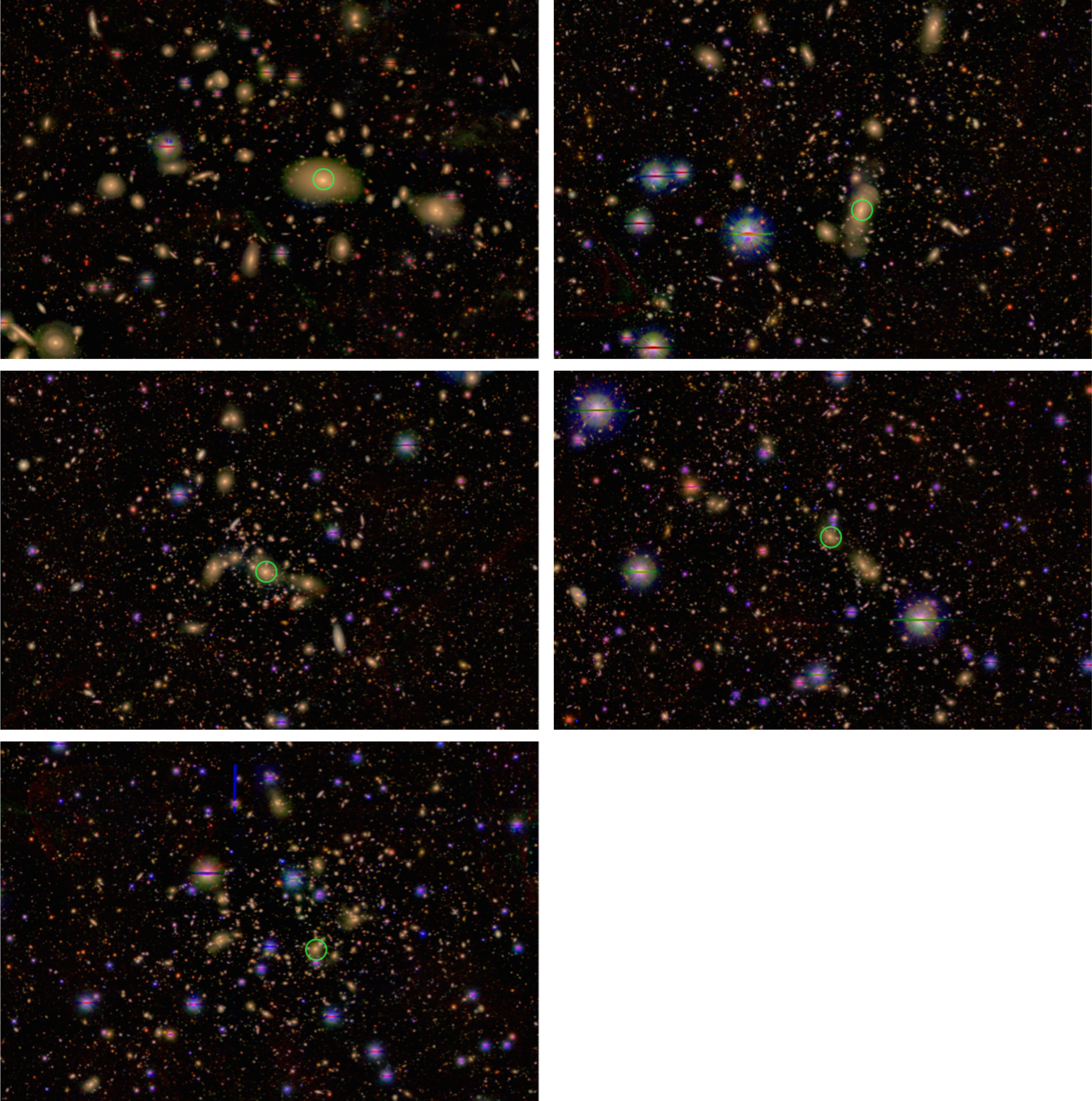}
\caption{Color ($riz$) images of {\it Planck }clusters observed by HSC Wide (top left to bottom left):  Abell~2457 ($z=0.06$), Abell~329 ($z=0.14$), Abell~362 ($z=0.18$), MaxBCGJ140.53188+03.76632 ($z=0.27$) and Abell~776 ($z=0.33$). Green circle marks the location of the BCG selected as the cluster center. The images are roughly $13\arcmin\times8.7\arcmin$ in scale.}
\label{fig:climages}
\end{figure*}

We make use of the {\it Planck } 2015 ``union" catalog \citep{Planck-Collaboration2016}, a union of all clusters detected by three  different {\it Planck } algorithms (\texttt{MMF1, MMF3, PwS}). We cross-match the {\it Planck } catalog with HSC FDFC footprint and find that five clusters are contained within the it. We attempt to maximize the lensing signal by stacking around the  brightest cluster galaxy (BCG) center, which provides a better trace of the center of potential than the SZ peak \citep{George2012} given the large {\it Planck} beam size. We visually inspect the HSC images of these five clusters in order to identify the BCG of each cluster. The five clusters, their BCG positions and redshifts are listed in Table~\ref{tab:clusters}. In Figure~\ref{fig:sky}, we show the sky distribution of all {\it Planck }clusters (gray points), the HSC-Wide FDFC fields that have been so far observed (gray regions), the planned HSC-Wide fields (black outline) and the {\it Planck } clusters detected within the current HSC S16A fields (circles colored by cluster redshift). In Figure~\ref{fig:climages}, we show color images of the five observed {\it Planck }cluster, acquired from the HSC imaging sky server tool, 
scMap\footnote{https://hsc-release.mtk.nao.ac.jp/hscMap/}. 
This figure indicates the clusters are often complex, non-relaxed and sometimes merging with other nearby groups.

\section{Weak Lensing Methodology}
\label{sec:WLmethod}
Weak lensing  distorts  the images of source galaxy shapes.  The amplitude of this distortion is proportional to all matter contained in the lensing cluster and along the line of sight to the lens.
The tangential distortion profile is related to the projected surface-mass density profile of the average mass distribution around the cluster,
\begin{equation} \label{eq:shear}
\gamma_T(R) = \frac{\DSigma(R)}{\Sigmacr} =  \frac{\bar{\Sigma}(<R)-\Sigma(R)}{\Sigmacr},
\end{equation}
where $R$ is the comoving transverse separation between the source and the lens, $\Sigma(R)$ is the projected surface mass density, $\bar{\Sigma}(<R)$ is the mean density within $R$, and
\begin{equation}\label{eq:sigcrit}
\Sigmacr =  \frac{c^2}{4\pi G}\frac{D_A(z_s)}{D_A(z_l)D_A(z_l,z_s)(1+z_l)^2},
\end{equation}
is the critical surface mass density, where $G$ is the gravitational constant, $c$ is the speed of light, $z_l$ and $z_s$ are the lens and source redshifts, respectively,  and $D_A(z_l)$, $D_A(z_s)$,   and $D_A(z_l,z_s)$ are the angular diameter distances to the lens, source, and between the lens and the source, respectively, and the extra factor of  $(1+z_l)^2$ comes from our use of comoving coordinates \citep{Bartelmann2001a}. 

We estimate the mean projected density contrast profile $\DSigma(R)$ from Equation~\ref{eq:shear} by stacking the shear over a population of source galaxies $s$ (over multiple clusters $l$) that lie within a given cluster-centric radial annulus $R$ (in comoving units),
\begin{equation}\label{eq:DSigma}
\DSigma(R) = \frac{1}{2\mathcal{R}(R)}\frac{ \sum\limits_{l,s} w_{ls} e_{t,ls}\big{[}\langle\Sigmacr^{-1}\rangle_{ls}\big{]}^{-1} } {(1+K(R)) \sum\limits_{l,s}w_{ls}},
\end{equation}
where the double summation is over all clusters and over all sources associated with each cluster (i.e., lens-source pairs), and
\begin{equation}\label{eq:et}
e_t= -e_1 \cos{2\phi} - e_2 \sin{2\phi},
\end{equation}
is the tangential shape distortion of a source galaxy, $\phi$ is the angle measured in sky coordinates from the right ascension direction to a line connecting the lens and source galaxy, and $e_1,e_2$ are the shear components in sky coordinates  obtained from the pipeline \citep{Mandelbaum2017,Bosch2017}. 
The mean critical density  $\langle\Sigma_{{\rm cr}}^{-1}\rangle_{ls}^{-1}$   is averaged with the source photo-z PDF, $P(z)$, for each lens-source pair, such that
\begin{equation}\label{eq:sigcritmean}
\langle\Sigmacr^{-1}\rangle_{ls} = \frac{ \int_{z_l}^{\infty} \Sigmacr^{-1}(z_l,z) P(z)\,\mathrm{d}z} { \int_{0}^{\infty} P(z)\,\mathrm{d}z}.
\end{equation}
As long as the mean $P(z)$ correctly describes the sample redshift distribution, the above 
equation  corrects for  dilution  by cluster or foreground source galaxies. However, obtaining realistic photo-z $P(z)$ is one of the biggest observational  challenges in WL analyses.
The weight in Equation~\ref{eq:DSigma}, $w_{ls}$, is given by
\begin{equation}
  \label{eq:lensweight}
w_{ls} = (\langle\Sigmacr^{-1}\rangle_{ls})^2 \frac{1}{\sigma_{e,s}^2+e_{{\rm rms},s}^2},
\end{equation}
where $\sigma_e$ is the per-component shape measurement uncertainty, and $e_{\rm rms}\approx0.40$ is the root mean square (RMS) ellipticity estimate per component.
The factor $(1 + K(R))$ corrects for a multiplicative shear bias $m$ as determined from the GREAT3-like simulations \citep{Mandelbaum2014,Mandelbaum2015} and is described in \cite{Mandelbaum2017a}. The factor is computed as
\begin{equation}
K(R) = \frac{\sum_{l,s}m_{s}w_{ls}}{\sum_{l,s}w_{ls}}.
\end{equation}
The  `shear responsivity' factor in Equation~\ref{eq:DSigma}, 
\begin{equation}
\mathcal{R}(R) = 1-\frac{\sum_{l,s} e^2_{\rm rms,s}w_{ls}}{\sum_{l,s} w_{ls}}\approx 0.84, 
\end{equation}
represents the response of the ellipticity, $e$, to a small shear \citep{Kaiser1995,Bernstein2002}. A full description and  clarification of this procedure is given in \cite{Mandelbaum2017}.

Finally, the bin-to-bin covariance matrix includes the statistical
uncertainty due to shape noise, the intrinsic variance of the
projected cluster lensing signal due to halo triaxiality and the
presence of correlated halos \citep{Gruen2015}, and cosmic-noise covariance due to uncorrelated  large-scale structures along the line-of-sight \citep{Hoekstra2003},
\begin{equation}
\mathsf{C} = \mathsf{C}^{\rm stat} + \mathsf{C}^{\rm int} + \mathsf{C}^{\rm lss}
\end{equation}
where
\begin{equation}
\mathsf{C}^{\rm stat}(R)=\frac{1}{4{\cal R}^2(R)}\frac{\sum_{l,s} {w}^2_{ls} (e^2_{{\rm rms},s} + \sigma_{e, s}^2)\left\langle \Sigma_{{\rm cr}}^{-1} \right\rangle_{ls}^{-2}}{\left[1+K(R)\right]^2\left[ \sum_{l,s} {w}_{ls}\right]^2}.
\end{equation}
The fractional intrinsic scatter is estimated to be 20\% of the
projected cluster lensing signal, per cluster, from semi-analytical
calculations calibrated by cosmological numerical simulations  \citep{Gruen2015,Umetsu2016,Becker2011}.
Following the prescription of \cite{Umetsu2016}, we assume the
diagonal form of the $\mathsf{C}^{\rm int}$ matrix, diag$[\mathsf{C}^{\rm int}(R)] =
[\alpha_\mathrm{int} \Delta\Sigma(R)]^2$, with
$\alpha_\mathrm{int}=0.2$. We do not expect significant bin-to-bin
covariance for the current binning scheme.
We compute the cosmic-noise covariance $C^{\rm lss}$ following \cite{Hoekstra2003}. We compute the elements of the $\mathsf{C}^{\rm lss}$ matrix using the
nonlinear matter power spectrum of \cite{Smith2003} for the
WMAP9 cosmology
\citep{Hinshaw2013}, with a source plane at $z=1.2$, the
mean redshift of our source galaxies (M17).
For the five Planck clusters, we simply scale the respective covariance matrices linearly according to the number of independent clusters: $\mathsf{C}\to \mathsf{C}/N$ with $N=5$. 

\section{Results and Analysis}
\label{sec:results}
In this section, we present the WL analysis of the five {\it Planck }clusters.  We show both the individual and stacked mass profiles  of the {\it Planck } clusters. We fit the cluster density profile with a model to obtain their total WL mass. Finally, we compare their WL masses with their measured {\it Planck }SZ mass and obtain the final SZ-WL mass calibration.

\subsection{Background Selection}
\label{subsec:CCsel}
We follow the methodology of \cite{Medezinski2010},  further explored and applied to HSC clusters in \cite[][hereafter M17]{Medezinski2017a}, in selecting background galaxies from the full galaxy sample. Two methods have been explored in M17 -- ``CC-cuts" \citep{Medezinski2010} and ``P-cut" \citep{Oguri2014}. CC-cuts relies on selection background galaxies in color-color (CC) space; specifically, for HSC the $g-i$ vs $r-z$ space has been used, where the cluster red-sequence can be well isolated in color from the background and foreground galaxies. The P-cut method relies on selecting galaxies whose photo-z PDF ($P(z)$) lie mostly beyond  the cluster redshift plus some threshold, i.e. $\int\limits_{z_l+{\Delta}z}^{\infty} P(z)\,\mathrm{d}z >0.98$. An optimized threshold is found to be  ${\Delta}z=0.2$.  M17 show that, with the above chosen limits,  these selections  provide consistent, undiluted WL profiles. Without these cuts the lensing signal is severely diluted for low-redshift clusters, as are the Planck-HSC clusters. We repeat this test here by utilizing the CC-cuts and P-cut source selections, and compare their profiles in the next section.

\subsection{Stacked Weak Lensing Analysis}
\label{subsec:WLstack}

\begin{figure}
\includegraphics[width=0.5\textwidth,clip]{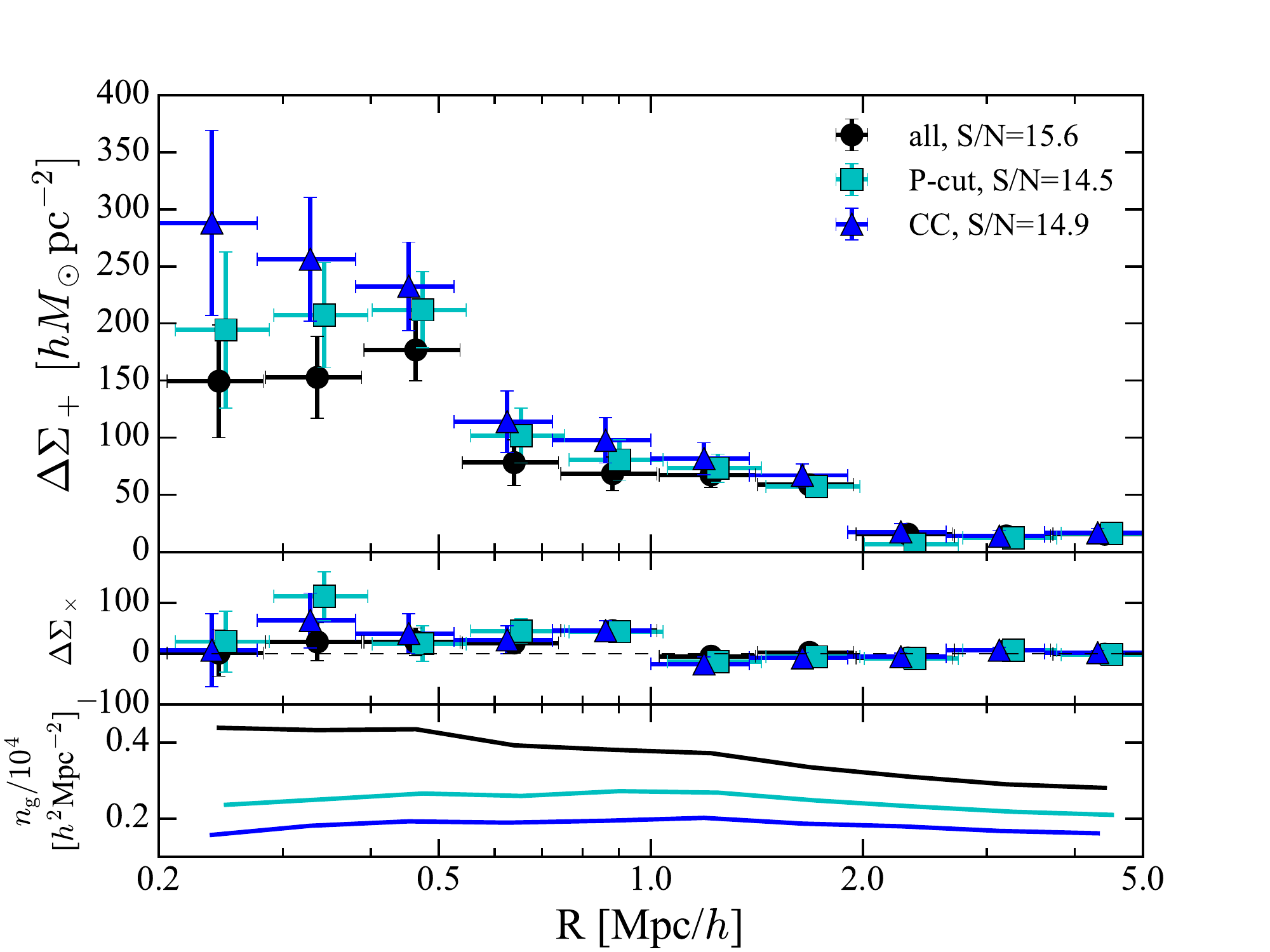}
\caption{Stacked surface mass density as a function of cluster-centric comoving radius (top panel). We compare between profiles derived using all galaxies (black circles), using only sources where their  $\sum P(z>z_{\rm cl}+0.2)>0.98$ (P-cut; cyan squares), and selecting only sources within color-color-magnitude cuts (CC-cuts; blue triangles). The mass density profile that uses all galaxies appears diluted inside the cluster  $\simlt0.7~\mpch$ relative to the more conservative P-cut/CC-cut profiles, mostly due to contamination from cluster members. The middle panels shows the $45\degree$-rotated shear, consistent with zero as expected.  The bottom panel shows the effective number density profile.}
\label{fig:DSigma_comp}
\end{figure}

We compute the mean lensing surface mass density profile, $\DSigma(R)$, given by Equation~\ref{eq:DSigma}, stacked over the five {\it Planck} cluster. We present the profiles in the top panel of Figure~\ref{fig:DSigma_comp} for the three selection methods: using the full sample (`all'; black circles), using the P-cut source sample (cyan squares), and using the CC-cut source sample (blue triangles). As can be seen from this comparison, the black points are systematically below the profiles of the other selection methods. However,  as opposed to the overall consistency between P-cut and CC-cuts found in M17, here the P-cut signal is systematically below the CC-cuts curve, though in agreement within the errors. We explore the impact of this dilution by comparing the fitted masses for the P-cut and CC-cut profiles next (Section~\ref{subsubsec:NFW}).

\subsection{NFW modeling}
\label{subsubsec:NFW}

To estimate the total mean mass of the clusters, we fit the stacked lensing profiles obtained in Section~\ref{subsec:WLstack}  with a universal  \citet*[][NFW]{NFW96}  mass density profile, given by the form 
\begin{equation}\label{eq:nfw}
\rho_{\rm NFW}(r)=\frac{\rho_{s}}{(r/r_s)(1+r/r_s)^{2}},
\end{equation} 
where $\rho_{s}$ is the characteristic density, and
$r_s$ is the characteristic scale radius at which the logarithmic density slope is isothermal. 
The halo mass $M_{\Delta}$ is  given by integrating the NFW profile (Equation~\ref{eq:nfw})
out to a  radius $r_{\Delta}$, at which the mean density is $\Delta\times\rho_\mathrm{crit}(z_\mathrm{l})$, the critical mass density of the universe at the cluster redshift, expressed as
$ M_{\Delta} \equiv  M(<r_{\Delta})= (4\pi/3)\rho_\mathrm{crit}(z_\mathrm{l})
      \Delta r_{\Delta}^3 $. We use $\Delta=200$ to define the halo
      mass, $ M_\mathrm{200c}$. The degree of concentration is defined as
      $c_\mathrm{200c}\equiv r_\mathrm{200c}/r_s$, and the characteristic density is
      then given by $\rho_s ={\Delta \rho_{\mathrm crit}}/{3} {c_\Delta^3}/[{\ln(1+c_\Delta)-c_\Delta/(1+c_\Delta)}]$.

The free parameters in this model are the mass,  $\mvir$, and concentration,  $\cvir$. We fix the mean cluster redshift to the lensing-weighted cluster redshift and fit for the mass and concentration using the Markov Chain Monte Carlo (MCMC) algorithm {\sc emcee} from \cite{Foreman-Mackey2014}. The 2-halo term becomes significant beyond 4 Mpc($\simeq3$~\mpch) for clusters in our mass range \citep{Oguri2011}, which coincides with our last two radial bins. We therefore exclude those bins and only model the 1-halo term. We also exclude the innermost radial bin, where masking  and imperfect deblending due to BCGs may affect our photo-z's or shape measurements (see discussion in M17; also R. Murata et al. in preparation). The final fitting range considered is 0.3--3~\mpch. For the sake of computational efficiency, we set  flat priors on the mass and concentration in the range $0\le M_\mathrm{200c}/(\mhunit)\le10$, $1\le c_\mathrm{200c}\le10$.

\begin{figure}
\includegraphics[width=0.5\textwidth,clip]{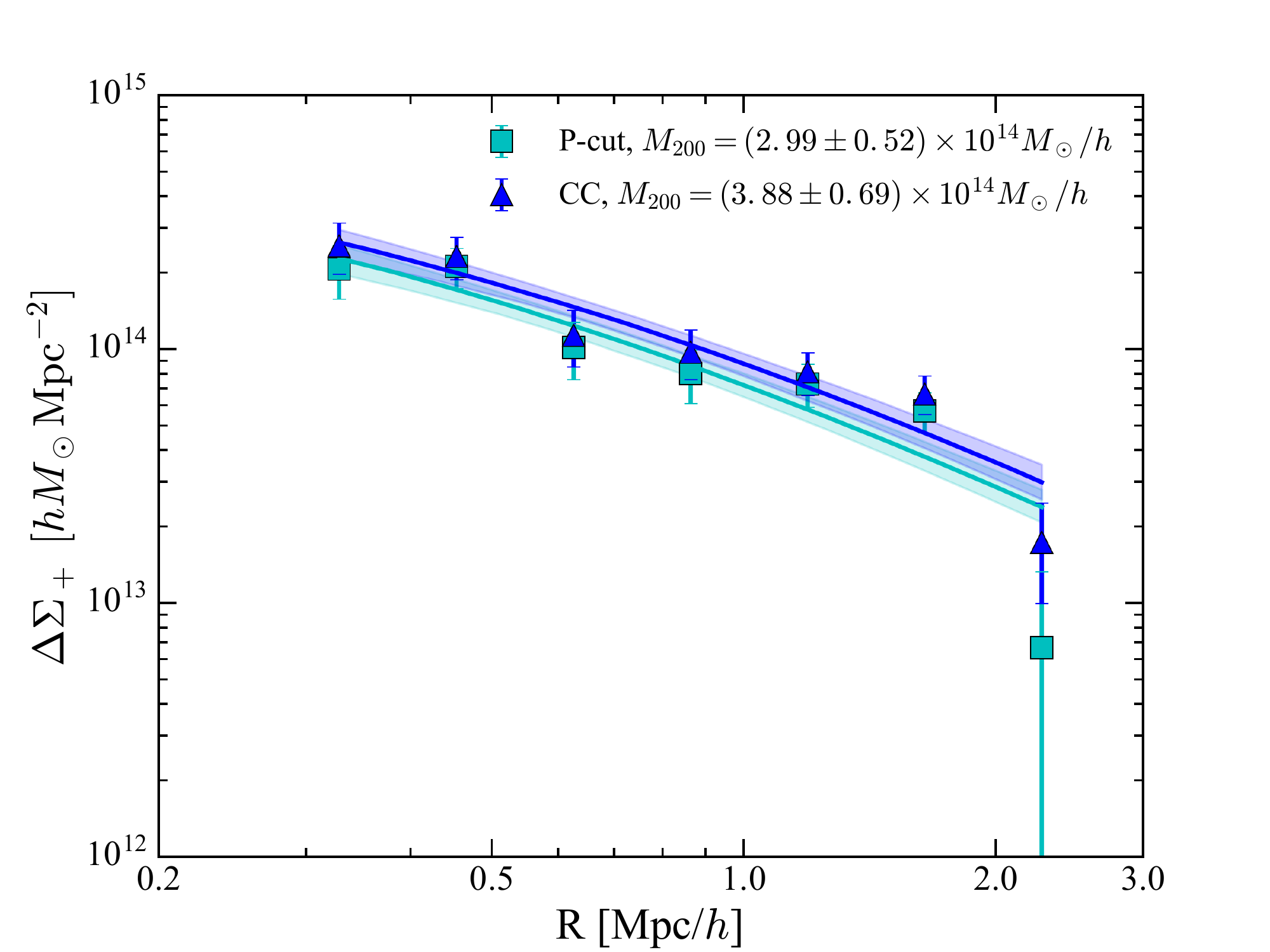}
\caption{NFW fit to the stacked surface mass density profile. 
Cyan squares and curve show the profile and its fit using P-cut galaxies behind the lens, and blue triangles and curve show the same for source galaxies selected with the CC-cut. The best-fit mass is given in the legend. Over 30\% bias is caused by contamination of cluster/foreground galaxies when not applying the CC-cuts. }
\label{fig:NFW}
\end{figure}

The profiles of the P-cut and CC-cut samples and their corresponding best-fit NFW profiles are shown in Figure~\ref{fig:NFW}  (cyan squares and blue triangles with fitted curves, respectively). The total masses, as derived from the median of the posterior distribution for each model parameter, are $\mvir = (3.88\pm0.69)\times\mhunit$ for the CC sample, and $\mvir = (2.99\pm0.52)\times\mhunit$ for the P-cut sample. It  therefore appears that the P-cut sample provides a mass estimate that is biased low by as much  as $\sim(23\pm19)\%$. Given that the P-cut profile is consistently below the CC profile at all radii, this either stems from foreground contamination or from faint cluster members. M17 show that the photo-z calibration bias is not large for low-z clusters, reaching at most $\sim4\%$.  
This discrepancy therefore cannot be fully explained by photo-z biases.
M17 also indicate that for massive (rich) low-redshift  clusters, cluster member dilution is more severe since we can probe fainter down the luminosity function. As it appears from Figure~\ref{fig:NFW}, these cluster contaminants are not adequately removed by the P-cut selection. The {\it Planck }clusters in HSC are all at low redshifts, $z<0.33$, and are fairly massive and rich -- the four clusters above $z>0.1$ have measured richness above 50 with either the SDSS/RedMaPPer  \citep{Rykoff2016} or the HSC/CAMIRA  \citep{Oguri2017} algorithms. 
As shown in M17 (see their Figure 6), a more stringent cluster redshift threshold could alleviate this contamination. We explore this here by setting $\Delta z=0.5$ for our P-cut sample, and find the fitted mass to somewhat increase, $\mvir = (3.46\pm0.68)\times\mhunit$, consistent with the mass found using the CC-cut sample. However, this significantly reduced the size of the P-cut sample to be even smaller than the CC-cut sample, so that it is no longer statistically preferred. 
We therefore hereon choose to work with the CC-cuts sample and use its derived mass. The mass and concentration fitted from the CC-cut profile are summarized in Table~\ref{tab:NFW}.

Uncertainties in the  identification of the cluster center may  bias the profile shape on small scales, and therefore the resulting mass estimate. Our visual inspection and selection of a BCG as the cluster center should mitigate this issue (compared with, e.g., using the SZ peak). However, as evident from Figure~\ref{fig:climages}, 
when significant substructure prevails in the center, the BCG determination may be ambiguous.
We do not have enough S/N in the inner radial bins to fully constrain the miscentering effect with a modified NFW model, but we may assess the level of uncertainty due to this effect. We  add a miscentered NFW model component convolved with a Rayleigh distribution \citep{Johnston2007},
\begin{equation}
\begin{aligned}
\Sigma& (R,M,c,R_{\rm mis},P_{\rm mis}) = \\
&( 1-P_{\rm mis} ) \Sigma(R,M,c) \\
&+ \frac{P_{\rm mis}}{2\pi} \int_{0}^{\infty} {\rm d} R' \big[ \frac{R'}{R_{\rm mis}^2} \exp{(\frac{-R'^2}{2R_{\rm mis}^2})}\\
&\times\int_{0}^{2\pi} {\rm d}\theta\Sigma(\sqrt{R^2+R'^2+2RR'\cos\theta},M,c) \big],
\end{aligned}
\end{equation}
but we fix the miscentering parameters and only fit for the mass and concentration. From $\Sigma(R,M,c,R_{\rm mis},P_{\rm mis})$ the density contrast $\DSigma(R,M,c,R_{\rm mis},P_{\rm mis})$ can then be readily computed. For the miscentering parameters, we use those derived in \cite{Oguri2017} who compared optically-selected CAMIRA cluster centers with their X-ray counterparts. They find that  the typical offset scale is $R_{\rm mis} = 0.26\pm0.04~\mpch$, and the  fraction of miscentered clusters is $P_{\rm mis} = 0.32\pm0.09$. We set flat prior around those parameters the size of the errors. 
We find  $\mvir = (4.21\pm0.71)\times\mhunit$ and $\cvir = 7.9\pm1.6$. Considering the mass difference between the models with and without miscentering, the systematic uncertainty is of the order $|3.88-4.21|/3.88\sim9\%$.

\begin{table*}
\tbl{NFW fitted mass, concentration and bias$^1$}{%
\begin{tabular}{lccccc}
\hline\hline
Name &  $M_{\rm WL,200c}$ & $c_{\rm 200c}$ & $M_{\rm WL,500c}$  & $M_{\rm SZ,500c}^2$ & $1-b$ \\
 & {[\mhunit]} &  & {[\munit]}  & {[\munit]} &  \\
 \hline
Abell2457 & $2.02^{+0.78}_{-0.66}$ &  & $2.11^{+0.78}_{-0.69}$ & $1.68^{+0.18}_{-0.20}$ & $0.80^{+0.39}_{-0.22}$ \\
Abell0329 & $2.21^{+1.17}_{-0.83}$ &  & $2.12^{+0.97}_{-0.79}$ & $3.74^{+0.33}_{-0.34}$ & $1.76^{+1.04}_{-0.55}$ \\
Abell0362 & $4.13^{+1.01}_{-0.90}$ &  & $4.50^{+1.05}_{-0.98}$ & $3.44^{+0.42}_{-0.43}$ & $0.77^{+0.21}_{-0.15}$ \\
MaxBCGJ140.53188+03.76632 & $31.03^{+21.24}_{-13.92}$ &  & $25.13^{+13.02}_{-9.33}$ & $5.03^{+0.44}_{-0.50}$ & $0.20^{+0.12}_{-0.07}$ \\
MACSJ0916.1-0023/Abell0776 & $8.10^{+3.26}_{-2.17}$ &  & $8.43^{+2.70}_{-2.10}$ & $4.23^{+0.55}_{-0.57}$ & $0.50^{+0.17}_{-0.12}$ \\\hline
Stacked &  $3.88\pm0.69$ & $6.3\pm1.8$ &  $4.15\pm0.61$  & $3.32\pm0.27$ & $0.80\pm0.14$ \\
\hline
\end{tabular} }
\label{tab:NFW}
\begin{tabnote}
$^1$ Using the source sample defined by CC-cuts.\\
$^2$ {\it Planck} SZ-derived masses, after 15\% Eddington bias correction (see Section~\ref{subsec:SZvWL}).
\end{tabnote}
\end{table*}

\subsection{Weak Lensing of individual clusters}
\label{subsec:WLsingle}

\begin{figure*}[t]
\includegraphics[width=\textwidth,clip]{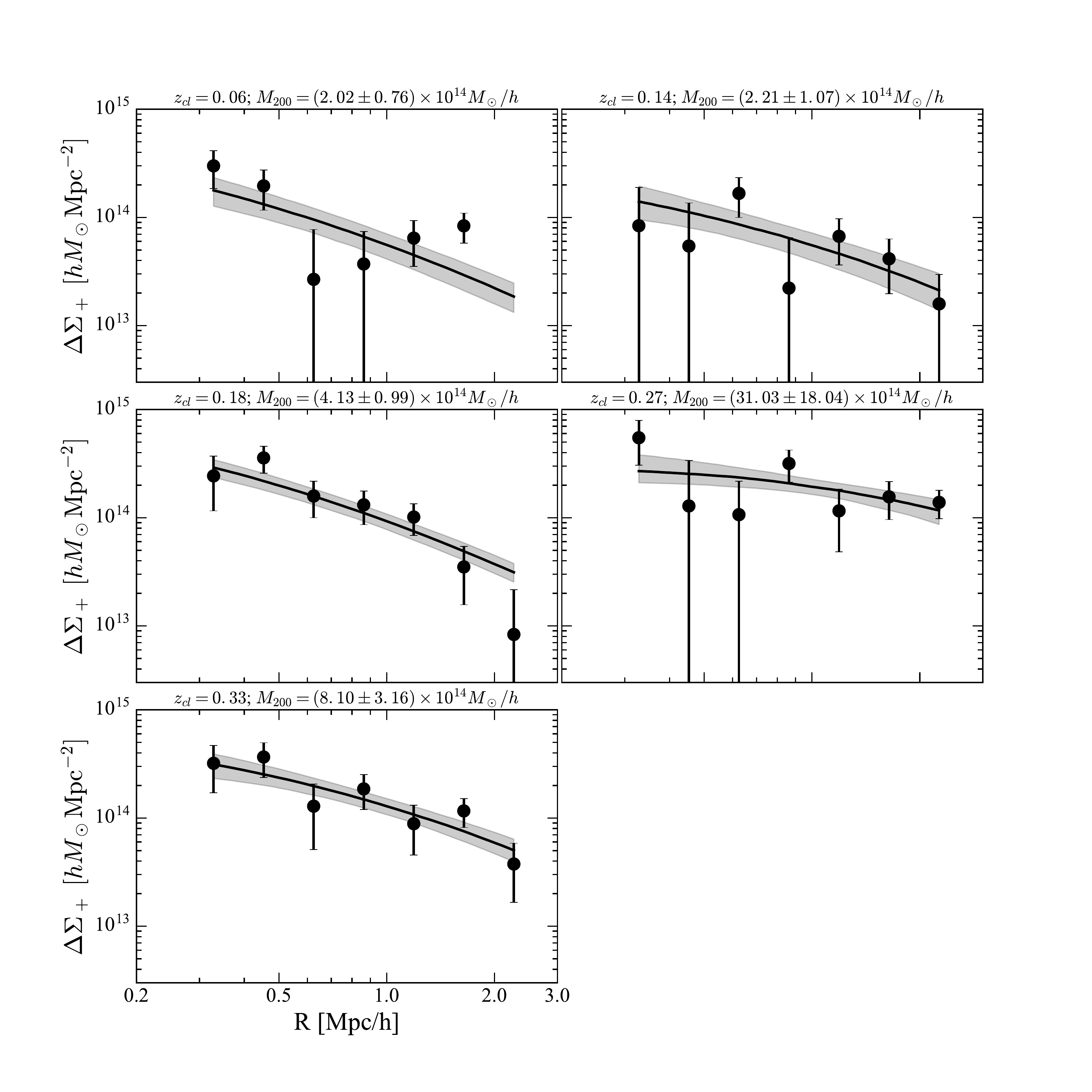}
\caption{Surface mass density profile for individual clusters. NFW profile fits are shown as black lines with 68\% confidence bounds. The cluster redshift and fitted mass are given for each cluster. }
\label{fig:DSigma_single}
\end{figure*}

\begin{figure}[t]
\includegraphics[width=0.5\textwidth,clip]{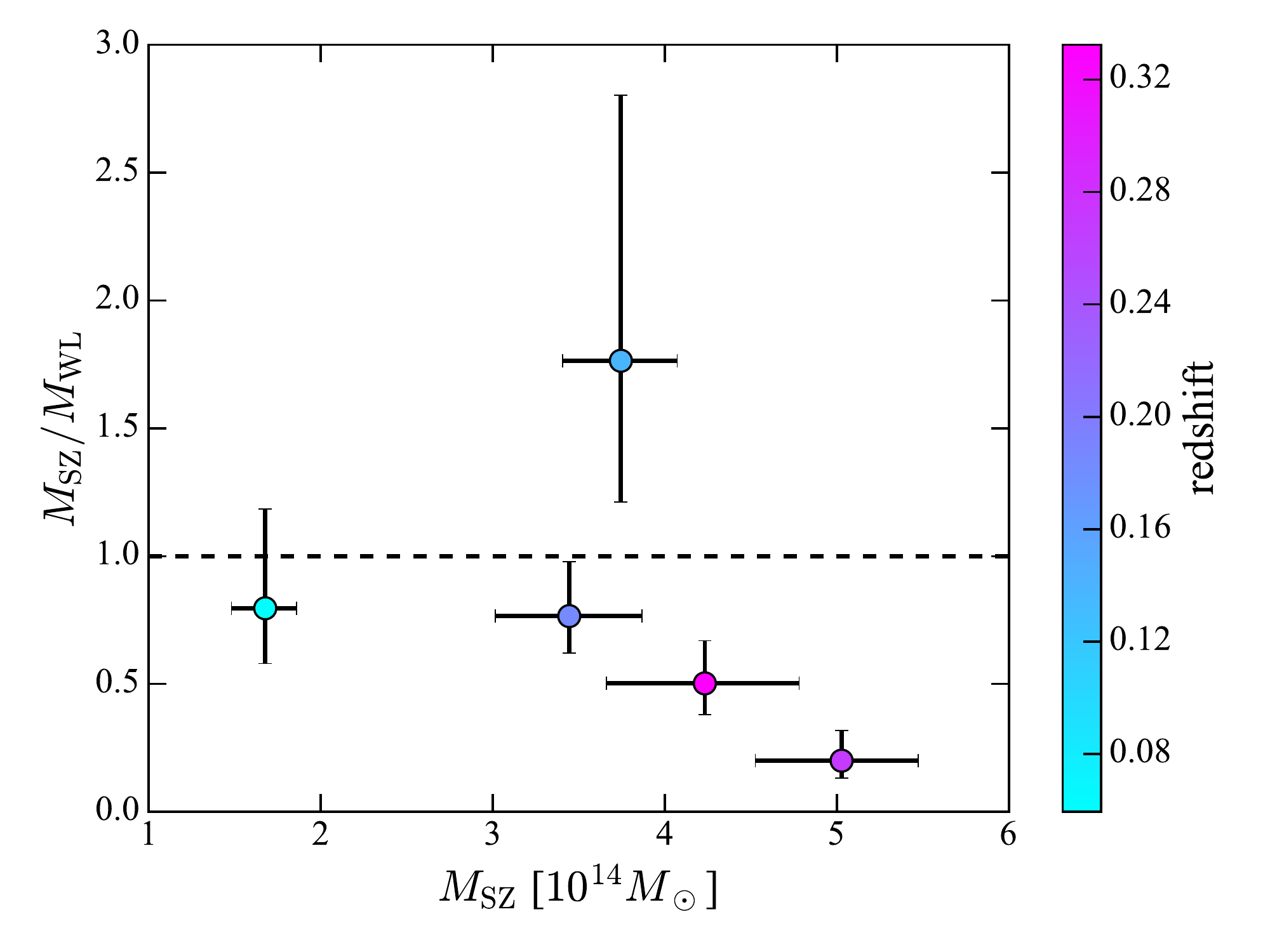}
\caption{ Ratio of the {\it Planck }SZ masses  to WL masses, $\msz/\mwl$, which depicts the level of bias as $1-b$, plotted as a function of $\msz$, for individual {\it Planck }clusters. The color scale represents cluster redshift. }
\label{fig:SZvWLsingle}
\end{figure}

Following the same procedure as above, we fit the  $\DSigma(R)$ individually for each cluster. 
Since we have lower S/N for each profile, we treat the concentration as a nuisance parameter in the range,  $1\le c_\mathrm{200c}\le10$, and allow for a broader range of masses, $0\le M_\mathrm{200c}/(\mhunit)\le100$.
We preset the  profiles and their fitted NFW  models (see Section~\ref{subsubsec:NFW}) in  Figure~\ref{fig:DSigma_single}. 
The fitted mass, along with the redshift, is given above each profile. We also translate the mass  to an overdensity of $\Delta=500$ to compare with the SZ values. We summarize the fitted WL masses and concentrations $M_{\rm 200c},\ c_{\rm 200c},\ M_{\rm 500c}$, the SZ masses, and the SZ-to-WL mass ratio in Table~\ref{tab:NFW} . All of our clusters have high S/N WL profiles, above S/N$\simgt6$. Here we define   S/N=$\sqrt{\sum_i{(\DSigma(R_i)/\mathrm{d}\DSigma(R_i))^2}}$, where our binning scheme ensures $\DSigma(R_i)/\mathrm{d}\DSigma(R_i)> 1$ in each radial bin per cluster, so that the
noise contribution to our S/N estimator is negligible. The highest S/N cluster is Abell\,362 at $z=0.18$, with S/N=10.4. This cluster is also present in the ACTPol SZ cluster catalog that is being used in an independent HSC  WL-SZ mass calibration study (Miyatake et al., in prep.). It is also present in the XMM-MCXC catalog and is being analyzed in an X-ray-WL mass calibration study by \citep{Miyaoka2017}. We all find consistent WL masses for this cluster. The most massive cluster, both in terms of SZ and WL mass (though with large errors), is MaxBCGJ140.53188+03.76632 at $z=0.27$. It is a relatively unstudied cluster, with a double-BCG disturbed morphology. For this cluster it was hard to determine which  BCG to use as a center, so we selected the one closest to the X-ray center, based on shallow (10~ksec) X-ray images from the {\it Chandra} archive (PI Rykoff).

Finally, we compare the SZ to WL mass by plotting  the ratio $\msz/\mwl$ as a function of SZ mass, color-coded by cluster redshift, in Figure~\ref{fig:SZvWLsingle}. Although there may appear to be a decreasing trend with increasing mass, our sample is small and the errors are large. This will be an interesting point to investigate with a future larger sample once the HSC completes the full Wide survey area.

\subsection{Systematics}
\label{subsec:syst}
Cluster WL analyses, and in particular when using observations as deep as HSC, may suffer several sources of systematic uncertainties.
As discussed here and thoroughly investigated  in M17, one of the main sources of systematics is due to  contamination from cluster members, and foreground galaxies whose photo-z's are not well represented by the PDF. Although we attempt to provide the most robust selection scheme to remove those from the source sample by applying the CC-cuts, some level of contamination may remain. To assess residual cluster contamination (if any), a boost factor is typically calculated. However, given the small sample of clusters studies here, this estimate will be unreliable, without availability of simulations (see discussion in M17). 

To assess foreground contamination robustly, a large,  spectroscopic sample  representative of HSC galaxies in terms of magnitude and colors is needed, which is not currently obtainable. The residual level of contamination estimated in M17 from a re-weighted spectroscopic redshift analysis appears minimal, $\lesssim4\%$ without any cuts, and $\lesssim2\%$ with the CC method used here.

As discussed in Section~\ref{subsubsec:NFW}, the miscentering of clusters can lead to 9\%  differences in the fitted mass, and  so we expect the systematic error due to miscentering to be of that order. 
To estimate the systematic error due to choice of radial range used in the modeling , we set  a more conservative inner radial cut, $R=0.5~\mpch$, similar to that used by WtG, and find $\mvir=(3.94\pm0.79)\times\mhunit$ for the CC-cut sample. This translates to a 1.5\% difference, which is small compared to the other sources of error.
In summary,  combined in quadrature, all  these errors  result in a 9\% systematic error,   below our statistical uncertainty (18\%).

\subsection{SZ Mass calibration}
\label{subsec:SZvWL}

Finally, we  address the level of bias between the {\it Planck } measured SZ cluster mass and that determined from the stacked lensing analysis presented in Section~\ref{subsubsec:NFW}. To do so, we first estimate the mean SZ mass of the five {\it Planck }clusters. We use the total lens-source weights (Equation~\ref{eq:lensweight}) for each lens to combine the SZ masses, such that the mean SZ mass is,
\begin{equation}\label{eq:meanSZ}
\langle\msz\rangle = \frac{1}{1+c_{EB}} \frac{\sum_{l}\msz_{,l} \sum_{s} w_{ls}}{\sum_{l,s} w_{ls}}
\end{equation}
where  the index $l$ runs over the five clusters, and the index $s$ identifies sources behind each cluster within the fitting range used in the model. In Equation~\ref{eq:meanSZ} we have applied an Eddington bias correction to the {\it Planck }SZ masses, $c_{EB} = 0.15$ . The correction here corresponds to the average difference in SZ masses found between ACT and
{\it Planck }\citep{Battaglia2016}, since ACT applied an Eddington
correction to their published SZ masses and {\it Planck } did not.
The resulting mean SZ cluster mass is $\langle\msz\rangle= (3.32\pm0.27)\times\munit$. To compare with the SZ mass, we convert the lensing mass fitted in Section~\ref{subsubsec:NFW} to the same overdensity, $\Delta=500$, and obtain $\langle\mwl\rangle = (4.15\pm0.61({\rm stat})\pm0.38({\rm sys}))\times\munit$.
Dividing the two masses,  we find the bias  for the five HSC-{\it Planck } clusters to be $1-b = \langle\msz\rangle/\langle\mwl\rangle = 0.80\pm0.14({\rm stat})\pm0.07({\rm sys})$.

We compare this value with those derived for the individual clusters in section~\ref{subsec:WLsingle}, by taking the {\em unweighted} mean of the ensemble following \cite{von-der-Linden2014}, so as not to be biased by the correlation of uncertainties with the $1-b$ values (lower $1-b$ values have lower errors). The mean ratio is $ \langle\msz/\mwl\rangle = 0.805$, which is in agreement with the stacked value above.

\begin{figure}[t]
\includegraphics[width=0.5\textwidth,clip]{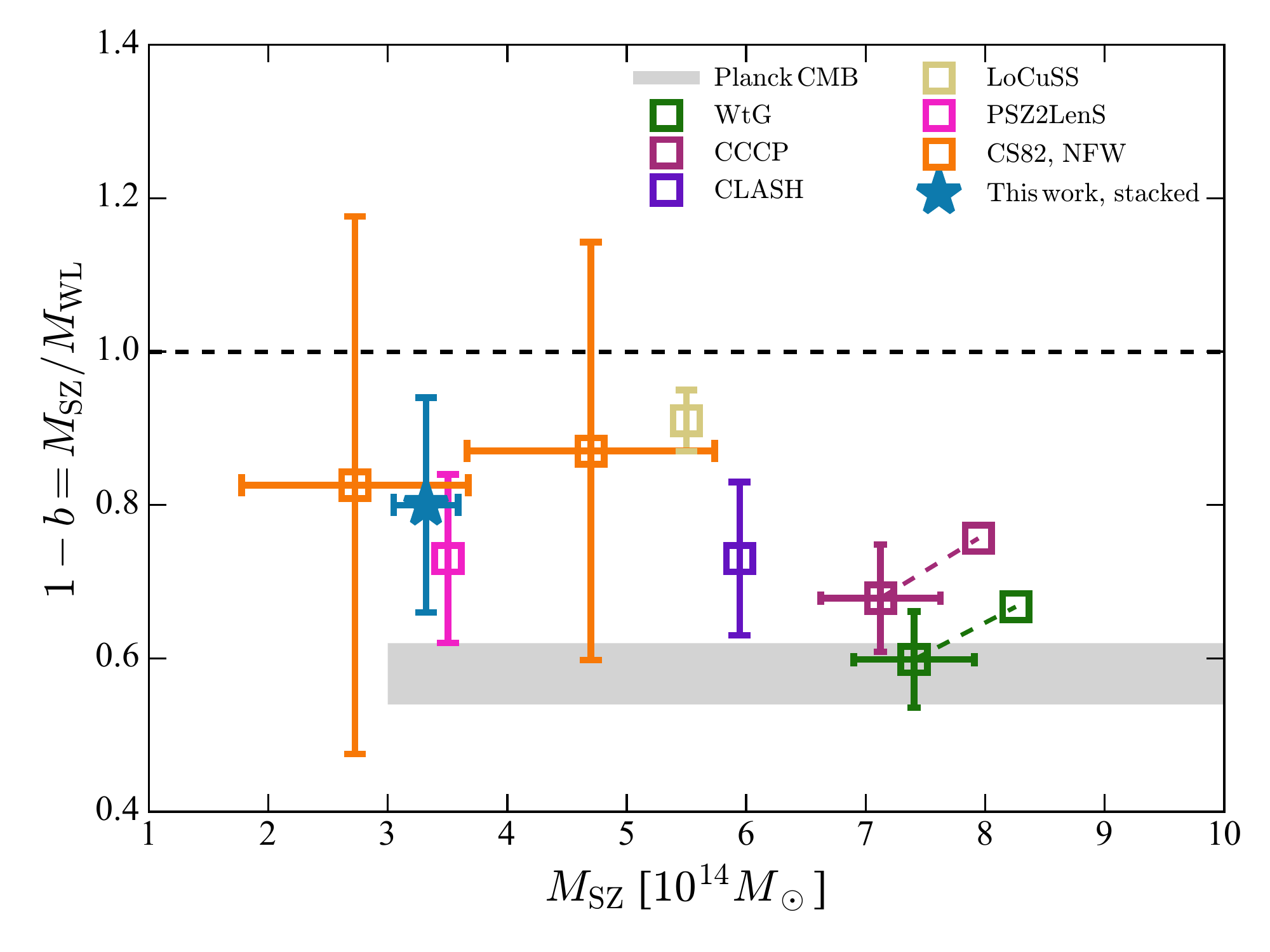}
\caption{ Ratio of the {\it Planck }SZ masses  to WL masses, $\msz/\mwl$, which depicts the level of bias as $1-b$, as a function of $\msz$. We compare the ratio inferred from the stacked lensing analysis done on the five {\it Planck }clusters in HSC (blue star) with different WL studies from the literature (squares) as indicated in the legend. All but the CS82 study are done for {\it Planck }clusters, as targeted by pointed WL observations. For WtG and CCCP, a range of Eddington bias is considered, 3\%--15\% (empty squares to squares with error bars, respectively). The comparison suggests that the level of bias, $1-b$, could be a function of cluster mass. }
\label{fig:SZvWLfinal}
\end{figure}

We present the stacked ratio in  Figure~\ref{fig:SZvWLfinal} (blue star) as a function of the mean SZ mass, and compare with other results from the literature. 
In the comparison, we consider a range of Eddington bias corrections, 3--15\% (dashed lines), for WL studies of {\it Planck} that did not apply this correction in their original analysis, namely WtG \citep[green squares;][]{von-der-Linden2014} and CCCP \citep[light-purple square;][]{Hoekstra2015}, as applied in \cite{Battaglia2016} (orange squares).
Our result is consistent, within the reported errorbars, to previous results over the same mass range in \msz~ (CS82 by \citealt{Battaglia2016} and PSZ2LenS by \citealt{Sereno2017a}). The values of $1-b$ found at higher  \msz~ differ by $\lesssim1\sigma$ (CLASH, \citealt{Penna-Lima2016} and CCCP, \citealt{Hoekstra2015}) to 2$\sigma$ (WtG, \citealt{von-der-Linden2014}), depending on the WL study and if considering the highest Eddington bias correction, 15\%. This reported difference as a function of \msz~ further supports the hypothesis that $1-b$ is a function of halo mass \citep[e.g.,][]{von-der-Linden2014,Hoekstra2015,Sereno2015}, although we cannot statistically conclude that a \msz~ dependence exists. With the five clusters all being at low redshift ($z<0.33$) we also cannot address the claims in \cite{Smith2016} that $1-b$ is a function of redshift \citep[see also][]{Sereno2017b}.

\section{Conclusions}
\label{sec:summary}

We have presented in this paper a WL analysis of five {\it Planck }clusters using the latest  $\sim140$~deg$^2$ deep multi-band HSC-SSP survey. We have taken steps to address different systematics that plague WL measurements. Using the  HSC photometry and shape measurement pipeline we correct for shape multiplicative bias. We  minimize foreground and cluster contamination of the source sample by applying CC-cuts. We measure the surface mass density profiles  both for the individual clusters and further stack them together to obtain a mean mass profile of $\sim15\sigma$. We fit the mass profiles with an NFW model, and find their mass range to be $(2$--$30)\times\munit$, and a mean mass of $\mwl_{,500c} = (4.15\pm0.61)\times\munit$. The level of mass bias with respect to the SZ mean mass is found to be  $1-b = \langle\msz\rangle/\langle\mwl\rangle = 0.80\pm0.14$. This low bias does not stand in tension to previous higher bias measurements, nor with the level needed to explain the high \sig8 found from primary {\it Planck }CMB, $1-b=0.58$, since we probe to a lower mass limit than previous studies. 
We note that the bias may be a function of cluster mass, however, we cannot conclude so based on this initial sample of only five clusters. To make more robust conclusions, we hope to revisit this analysis with a future larger sample of clusters.

When the full HSC-Wide survey is complete in 2019, it will have observed $\sim1400$ deg$^2$, with which we expect to have observe 10 times more {\it Planck} clusters. The level of uncertainty on the mass calibration, if assuming it is statistics dominated and scales as $N^{-1/2}$, will be reduced from the current $10\%$ for the five {\it Planck}-HSC clusters to reach $\sim3\%$ using 50 clusters. This level will be below what we currently find the systematic uncertainty to be, $\sim 9\%$, and will therefore require an even more robust treatment of cluster contamination, improvement to photo-z codes, and modeling. With such a high S/N measurement ($\sim50\sigma$ expected), we will be able to provide a tighter mass calibration and re-derive cosmological constraints from {\it Planck} SZ cluster counts below the current $10\%$ level. We will further gain insight on the mass bias due to the  HSE  assumption and study its possible dependence on cluster mass and redshift, informing us about the evolution of clusters and their gas physics.

\begin{ack}
EM acknowledges fruitful discussions with Andy Goulding and Peter Melchior.
The Hyper Suprime-Cam (HSC) collaboration includes the astronomical communities of Japan and Taiwan, and Princeton University. The HSC instrumentation and software were developed by the National Astronomical Observatory of Japan (NAOJ), the Kavli Institute for the Physics and Mathematics of the Universe (Kavli IPMU), the University of Tokyo, the High Energy Accelerator Research Organization (KEK), the Academia Sinica Institute for Astronomy and Astrophysics in Taiwan (ASIAA), and Princeton University. Funding was contributed by the FIRST program from Japanese Cabinet Office, the Ministry of Education, Culture, Sports, Science and Technology (MEXT), the Japan Society for the Promotion of Science (JSPS), Japan Science and Technology Agency (JST), the Toray Science Foundation, NAOJ, Kavli IPMU, KEK, ASIAA, and Princeton University. 
This paper makes use of software developed for the Large Synoptic Survey Telescope. We thank the LSST Project for making their code available as free software at  http://dm.lsst.org.
The Pan-STARRS1 Surveys (PS1) have been made possible through contributions of the Institute for Astronomy, the University of Hawaii, the Pan-STARRS Project Office, the Max-Planck Society and its participating institutes, the Max Planck Institute for Astronomy, Heidelberg and the Max Planck Institute for Extraterrestrial Physics, Garching, The Johns Hopkins University, Durham University, the University of Edinburgh, Queen?s University Belfast, the Harvard-Smithsonian Center for Astrophysics, the Las Cumbres Observatory Global Telescope Network Incorporated, the National Central University of Taiwan, the Space Telescope Science Institute, the National Aeronautics and Space Administration under Grant No. NNX08AR22G issued through the Planetary Science Division of the NASA Science Mission Directorate, the National Science Foundation under Grant No. AST-1238877, the University of Maryland, and Eotvos Lorand University (ELTE) and the Los Alamos National Laboratory.
Based (in part) on data collected at the Subaru Telescope and retrieved from the HSC data archive system, which is operated by Subaru Telescope and Astronomy Data Center at National Astronomical Observatory of Japan.
This paper makes use of packages available in Python's open scientific
ecosystem, including NumPy \citep{NumPy}, SciPy
\citep{SciPy},  matplotlib
\citep{matplotlib}, IPython \citep{IPython},
AstroPy \citep{astropy13}, and cluster-lensing \citep{Ford2016}.
The work reported on in this paper was substantially performed at the TIGRESS high performance computer center at Princeton University which is jointly supported by the Princeton Institute for Computational Science and Engineering and the Princeton University Office of Information Technology's Research Computing department.
NB acknowledges the support from the Lyman Spitzer Jr. Fellowship.
HM is supported by the Jet Propulsion Laboratory, California Institute of Technology, under a contract with the National Aeronautics and Space Administration.
This work was supported in part by World Premier 
International Research Center Initiative (WPI Initiative), 
MEXT, Japan, and JSPS KAKENHI Grant Number 
26800093 and 15H05892.
KU acknowledges support from the Ministry of Science and Technology of Taiwan
through the grant MOST 103-2112-M-001-030-MY3.

\end{ack}

\bibliography{/Users/elinor/Dropbox/Documents/Elinor.bib}

\end{document}